\documentstyle[twocolumn,prb,aps]{revtex}
\renewcommand{\narrowtext}{\begin{multicols}{2} \global\columnwidth20.5pc}

\begin{document}
\newcommand{\be}{\begin{equation}}
\newcommand{\ee}{\end{equation}}
\newcommand{\nt}{\narrowtext}

\noindent
Comment on $\bf Electron$ $\bf Spectral$ $\bf Function$
$\bf and$\\ $\bf Algebraic$ $\bf Spin$ $\bf Liquid$ $\bf for$ 
$\bf the$ $\bf Normal$ $\bf State$ $\bf of$ \\
$\bf Underdoped$ $\bf High$ $\bf T_c$ $\bf Superconductors$ 

\noindent
In a recent Letter [1], W. Rantner and X.-G. Wen made a 
theoretical prediction of the power-law behavior of the electron spectral function 
in the pseudogap phase
of underdoped cuprates, reminiscent of that in the one-dimensional Luttinger liquid. 
This conclusion was drawn on the basis
of the following (somewhat heuristic) propositions: 1) the properties of 
the pseudogap phase are described by the slave-boson $QED_3$-like effective theory
formulated in terms of neutral spinons $\psi$, charged holons,
and a gauge field $A_\mu$;
2) the propagator of physical electrons $G_e$ can be computed as a simple product of the 
spinon ($G_s$) and the holon ($G_h$) ones; 3) provided that the holons are (nearly) condensed 
($G_h\approx const$), $G_e$ becomes proportional to the gauge-invariant spinon amplitude
(here $x_\mu=(t, {\bf r})$ is a position vector in the three-dimensional space-time) 
\be
G_s(x)=<\psi(x)\exp(-i\int_\Gamma A_\mu(z)dz^\mu){\overline \psi}(0)> 
\ee
with the contour $\Gamma$ chosen as a straight line between the end points;
4) the amplitude (1) decays algebraically,
$G_s(x)\sim {1/|x|^{2+\eta}}$, and the anomalous exponent $\eta$ is $positive$.

It turns out, however, that the value of $\eta$ quoted in Ref.[1]
and later derived in the original (the only one available at the time of submitting this Comment) 
version of Ref.[2] had a wrong sign, as was 
pointed out in Ref.[3] where instead a $negative$ value, 
$\eta=-32/3\pi^2N$, was obtained (the number $N$ of fermion species in cuprates is $N=2$).

In fact, thus far no physically motivated gauge-invariant alternative to Eq.(1) 
that would exhibit a power-law decay 
with a positive $\eta$ has ever been constructed, and it remains unknown if
such a function can exist at all in the pure massless $QED_3$.
In turn, the negative value of $\eta$ disqualifies Eq.(1) introduced in Ref.[1] 
from being a viable candidate to the 
role of the gauge-invariant spinon propagator
(let alone the physical electron one), because
instead of the anticipated $suppression$ (as in other examples of doped Mott insulators
governed by strong electron correlations), 
the amplitude (1) manifests $enhancement$ as compared to the mean-field
($N=\infty$) result.

The argument appears to be particularly compelling in the 
limit of zero doping, in which case,
while preventing the electrons' spatial motion by making 
$G_e(x)$ vanish for any ${\bf r}\neq {\bf 0}$, the holon factor
$G_h(x)$ does not affect the amplitude $G_e(t,{\bf 0})$ which is directly related to the $physical$
electron density of states (DOS) proportional to $Im\int dt e^{i\epsilon t}G_e(t, {\bf 0})\sim
|\epsilon|^{1+\eta}$. Therefore, a negative $\eta$ would have given rise to a sub-linear DOS 
which is increased with respect to the mean-field "V-shaped" one.

Thus, taken at its face value, the negative $\eta$ invalidates 
the main prediction of Ref.[1] regarding the Luttinger-like
behavior of the electron spectral function $A(\epsilon, {\bf p})\sim Im G_e(\epsilon, {\bf p})$ 
which, under the above assumptions, was identified in Ref.[1] with the Fourier transform
of Eq.(1). 
Moreover, once the holon factor $G_h(x)$ becomes non-trivial as well, 
the electron spectral function
(now given by a convolution of the Fourier transforms of $G_s(x)$ and $G_h(x)$)
can no longer feature a simple algebraic behavior, unless $both$ functions decay as power-law.
In the absence of any evidence suggesting otherwise, however, 
the possibility of such a behavior for the holons seems to be even more 
remote than for the spinons.

Furthermore, unless proven wrong, the absence of a physically sensible alternative 
to Eq.(1) in massless $QED_3$ may indicate a need for a revision of 
some of the above propositions which the work of Ref.[1] was based upon.
Indeed, albeit obtained in the framework of a perturbative $1/N$-expansion,
the unphysical behavior of $A(\epsilon, {\bf p})$ derived from 
Eq.(1) would have manifested itself at all energies/temperatures above a 
characteristic scale associated with such $non$-$perturbative$ effects as spinon 
chiral symmetry breaking, holon condensation, and/or gauge field instantons.
At still lower energies/temperatures, however, any of the above mechanisms may generate
a finite spinon and/or gauge field gap, thereby 
drastically altering the power-law decay of Eq.(1), consistent with the anticipated 
onset of such intrinsic instabilities of the pseudogap 
phase as antiferromagnetism, superconductivity and/or stripe order.

To conclude, despite its strong intellectual appeal, the $QED_3$-theory of 
underdoped cuprates has not yet provided
a firm theoretical support for the Letter's prediction of the Luttinger-like or $algebraic$
(which must be distinguished from both a $generic$ non-Fermi liquid, characterized by a 
mere absence of the quasiparticle peak, and a $virtually$ spin-charge separated
Fermi liquid which has a small coherent peak at low energies) behavior of the
electron spectral function, thus still leaving unsubstantiated the claim
of its satisfactory agreement with the normal state photoemission data 
which was made in Ref.[1] on the basis of the original erroneous evaluation of Eq.(1).

This research was supported by the NSF under Grant DMR-0071362.

$\bf D. V. Khveshchenko$\\
Department of Physics and Astronomy, 
University of North
Carolina, Chapel Hill, NC 27599

\end{document}